\documentclass[draftcls, 12pt,onecolumn,oneside]{IEEEtran}

\usepackage{times,amsmath,amssymb,graphicx,epsfig,cite,geometry,color,psfrag}
\makeatletter
\let\IEEEproof\proof
\let\IEEEendproof\endproof
\let\proof\@undefined
\let\endproof\@undefined
\makeatother

\usepackage{epsfig}
\usepackage{graphicx}
\usepackage{amsmath}
\usepackage{amsthm}
\usepackage{psfrag}
\usepackage{cite}
\usepackage{url}
\usepackage{amsfonts}
\usepackage{amssymb}

\let\proof\IEEEproof
\let\endproof\IEEEendproof

\allowdisplaybreaks
\newcommand{\bit}{\begin{itemize}}
\newcommand{\eit}{\end{itemize}}
\newcommand{\ben}{\begin{eqnarray*}}
\newcommand{\een}{\end{eqnarray*}}
\newtheorem {theorem}{Theorem}
\newtheorem {lemma}{Lemma}
\newtheorem {example}{Example}

\newtheorem{algorithm}[theorem]{\bf Algorithm}
\def\GF{{\rm GF}}

\begin{document}
\bibliographystyle{ieeetran}
\title{Retrieving Reed-Solomon coded data under interpolation-based list decoding}
\author{Jianwen~Zhang,~\IEEEmembership{Student Member,~IEEE} \\Marc A.~Armand,~\IEEEmembership{Member,~IEEE}
\thanks{Jianwen ~Zhang, and Marc A.~Armand are  with the Department of Electrical $\&$ Computer
Engineering, National University of Singapore, 119260.
(Email:\{jianwen.zhang, elema\}@nus.edu.sg)}} \maketitle
\begin{abstract}
A transform that enables generator-matrix-based Reed-Solomon (RS)
coded data to be recovered under interpolation-based list decoding
is presented. The transform matrix needs to be computed only once
and the transformation of an element from the output list to the
desired RS coded data block incurs $k^{2}$ field multiplications,
given a code of dimension $k$.
\end{abstract}
\begin{keywords}
 Galois Field Fourier Transform, list decoding, Reed-Solomon codes
\end{keywords}
\section{INTRODUCTION}\label{sectintr}
Since their inception in 1958, Reed-Solomon (RS) codes have found widespread use, e.g., in compact discs (CDs), digital video broadcasting and high definition TV. Let $\Phi$ be a cyclic subgroup of $\GF(q)\backslash\{0\}$ of order $n$ and $\alpha$ its generator. Then, following its original definition, $\{(c_{0},c_{1},\dots,c_{n-1})\in\GF(q)^{n}: c_{i}=m(\alpha^{i})\mbox{, } 0\leq i\leq n-1\mbox{, }  m(x)=\sum_{j=0}^{k-1}m_{j}x^{j}\in\GF(q)[x]\}$ is an $(n,k)$ RS code over $\GF(q)$ with zeros $\alpha,\alpha^{2},\ldots,\alpha^{n-k}$. In practice however, data is typically not encoded via the evaluation map
\begin{eqnarray} \label{TheEvaluationMap}
E_{\Phi} & : & \GF(q)^{k}\mapsto \GF(q)^{n} \nonumber \\
& : & (m_{0},m_{1},\ldots,m_{k-1}) \rightarrow (m(1),m(\alpha),\ldots,m(\alpha^{n-1}))
\end{eqnarray}
implied by this definition. On the other hand, the
interpolation-based RS list decoders of
\cite{sudan:guru,kotter:vardy} which allow far more errors to be
corrected than previously thought possible, operate on the
assumption that messages are encoded via an evaluation map.
Incorporating such a decoder within an existing system employing
RS codes is clearly desirable. For instance, it will allow a
system to operate at a significantly lower signal-to-noise ratio
while maintaining the same error performance.
Nevertheless, incorporating such a decoder into an existing system
requires the ability to recover the original RS coded data from
the output list of this decoder. For the CD player for example,
this would mean that we will not have to replace all our existing
CDs. A method for recovering generator-matrix-based RS coded data under
interpolation-based list decoding is therefore desirable as well.

Such a method has implicitly been introduced in the proof of
\cite[Lemma 4]{guruswami:Vardy}. The main idea there is as
follows. Let $v_{0},v_{1},\ldots,v_{n-1}$ be nonzero elements of
$\GF(q)$. Then the codeword $(c_{0},c_{1},\dots,c_{n-1})$ of an
$(n,k)$ generalized RS code defined by a basis of codewords given
by the rows of the matrix
\begin{equation}\label{Gp}
    {\cal{G}}=\left(\begin{array}{cccc}
v_{0} & v_{1} & \ldots & v_{n-1} \\
v_{0} & v_{1}\alpha & \ldots & v_{n-1}\alpha^{n-1} \\
\vdots & \vdots & \ddots & \vdots \\
v_{0} & v_{1}\alpha^{k-1} & \ldots & v_{n-1}\alpha^{(k-1)(n-1)}
\end{array}\right)
\end{equation}
and the codeword $(c_{0}/v_{0},c_{1}/v_{1},\dots,c_{n-1}/v_{n-1})$
of an $(n,k)$ RS code defined by a different basis of codewords
given by the rows of the matrix
$$\left(\begin{array}{cccc}
1 & 1 & \ldots & 1 \\
1 & \alpha & \ldots & \alpha^{n-1} \\
\vdots & \vdots & \ddots & \vdots \\
1 & \alpha^{k-1} & \ldots & \alpha^{(k-1)(n-1)}
\end{array}\right)$$
correspond to the same message. Thus, suppose the message
$(m_{0},m_{1},\ldots,m_{k-1})$ is encoded via multiplication by
$\cal{G}$ and the resulting codeword transmitted and received as
$(r_{0},r_{1},\ldots,r_{n-1})$. Provided less than
$n-\sqrt{(k-1)n}$ errors occur during transmission, the decoder of
\cite{sudan:guru}, operating on the premise that messages are
encoded via the evaluation map in (\ref{TheEvaluationMap}),
generates a list containing $(m_{0},m_{1},\ldots,m_{k-1})$ when
applied to the modified received word
$(r_{0}/v_{0},r_{1}/v_{1},\ldots,r_{n-1}/v_{n-1})$. The underlying
basis transformation thus results in an overhead of $n$
multiplications to the overall decoding complexity. The above
method is however no longer applicable when the generator matrix
$\cal{G}$ is not of the form in (\ref{Gp}), e.g. when $\cal{G}$
has the same structure as the matrix in (\ref{generatorg}).


This correspondence presents a more general solution to the
aforementioned problem in that our technique remains applicable
given {\em any} generator matrix $\mathbf{G}_a$ for an RS code.
Since $\mathbf{G}_a$ is arbitrary, the underlying basis
transformation employed in our approach is different to that used
in \cite{guruswami:Vardy} - see Lemma \ref{l1} below. In fact, if
messages are encoded as codewords of a narrow-sense RS code, {\em
no} basis transformation is needed. For code rates of practical
interest, an average computational overhead of $O(k^2)$
multiplications is added to the decoding process.

We exploit certain properties of the Galois Field Fourier
Transform (GFFT), including the fact that encoding a message block
$(m_{0},m_{1},\ldots,m_{k-1})$ via the evaluation map in
(\ref{TheEvaluationMap}) is equivalent to computing the GFFT of
the $n$-tuple $(m_{0},m_{1},\ldots,m_{k-1},0,\ldots,0)$ -- a
consequence of property 2 of \cite[Theorem 8-13]{Stephen:code}.
Our main result is Theorem \ref{t1} in Section \ref{sectkey}
below. However, before we can prove this theorem, we need a few
lemmas. For the remainder of this correspondence, take the
evaluation map to mean that specified by (\ref{TheEvaluationMap}).
Further, take an interpolation-based list decoder to mean an
instance of the decoders of \cite{sudan:guru,kotter:vardy}.

\section{LEMMAS LEADING TO THE KEY RESULT}\label{sectlemma}
Without loss of generality, let $q$ be a fixed power of $2$.
Suppose $\mathcal{C}$ is an ($n,k$) RS code over $\GF(q)$ with
generator polynomial
$g(x)=\prod^{n-k-1+b}_{i=b}(x-\alpha^{i})=\sum^{n-k}_{i=0}g_{i}x^{i}$.
We do not assume that $b=1$ so that $\mathcal{C}$ is not
necessarily a narrow-sense RS code.
\begin{lemma}\label{l1}
Let $$\mathbf{W}=\left(%
\begin{array}{cccc}
  1 & 0 & \cdots & 0 \\
  0 & \alpha^{(b-1)} & \cdots & 0 \\
  \vdots & \vdots & \ddots & \vdots \\
  0 & 0 & \cdots & \alpha^{(b-1)(n-1)} \\
\end{array}%
\right).$$ If $\mathbf{c}=(c_{0},c_{1},\ldots,c_{n-1})\in
\mathcal{C}$, then $\bar{\mathbf{c}}=\mathbf{c}\times \mathbf{W}$
is a codeword of a narrow-sense ($n,k$) RS code
$\bar{\mathcal{C}}$ over $\GF(q)$.
\end{lemma}
\begin{proof}
Let the code polynomial for $\mathbf{c}$ and $\bar{\mathbf{c}}$ be
written as $c(x)=\sum^{n-1}_{i=0}c_{i}x^{i}$ and $\bar{c}(y)$.
Then
\begin{equation}\label{modify}
    c(x)=\sum^{n-1}_{i=0}c_{i}\alpha^{(b-1)i}\left(\frac{x}{\alpha^{b-1}}\right)^{i}=\sum^{n-1}_{i=0}\bar{c}_{i}y^{i}=\bar{c}(y),
\end{equation}
where $\bar{c}_{i}=c_{i}\alpha^{(b-1)i}$,
$y=\frac{x}{\alpha^{b-1}}$. Since $c(x)$ has zeros
$\alpha^{b},\alpha^{b+1},\ldots,\alpha^{n-k-1+b}$, $\bar{c}(y)$
clearly has zeros $\alpha,\alpha^{2},\ldots,\alpha^{n-k}$ and the
lemma follows.
\end{proof}

Next, let $\bar{g}(y)=\sum^{n-k}_{i=0}\bar{g}_{i}y^{i}$ where
$\bar{g}_{i}=g_{i}\alpha^{(b-1)i}$. Since $\bar{g}(y)$ may not be
monic, it is a code polynomial, but not necessarily, the generator
polynomial of $\bar{\mathcal{C}}$. The following matrix is
nevertheless
\begin{equation}\label{barg}
\mathbf{\bar{G}}=\left(\!\!\!%
\begin{array}{ccccccc}
  \bar{g}_{0} & \bar{g}_{1} & \cdots & \bar{g}_{n-k} & 0 & \cdots & 0 \\
  0 & \bar{g}_{0} & \cdots & \bar{g}_{n-k-1} & \bar{g}_{n-k} & \cdots & 0 \\
  \vdots & \ddots & \ddots & \ddots & \ddots & \ddots & \vdots \\
  0 & 0 & \cdots & \cdots & \cdots & \cdots & \bar{g}_{n-k}
\end{array}%
\!\!\!\right).
\end{equation}
a valid generator matrix for $\bar{\mathcal{C}}$, since the rows
of $\bar{\mathbf{G}}$ spam a vector space over $\GF(q)$ of
dimension $k$. Let $[\mathbf{U}]$ denote the $n\times n$ matrix
resulting from appending $n-k$ rows to a $k\times n$ matrix
($n>k$) $\mathbf{U}$ such that each additional row is a right
cyclic shift of the previous row by one position. Lemma \ref{l2}
shows the relation between $[\bar{\mathbf{G}}]$ and $[\mathbf{G}]$
where
\begin{equation}\label{generatorg}
\mathbf{G}=\left(\!\!\!%
\begin{array}{ccccccc}
  g_{0} & g_{1} & \cdots & g_{n-k} & 0 & \cdots & 0 \\
  0 & g_{0} & \cdots & g_{n-k-1} & g_{n-k} & \cdots & 0 \\
  \vdots & \ddots & \ddots & \ddots & \ddots & \ddots & \vdots \\
  0 & 0 & \cdots & \cdots & \cdots & \cdots & g_{n-k}
\end{array}%
\!\!\!\right)
\end{equation}
is a generator matrix for the code $\mathcal{C}$.
\begin{lemma}\label{l2}
$[\mathbf{G}]\times \mathbf{W}=\mathbf{W}\times
[\bar{\mathbf{G}}]$.
\end{lemma}
\begin{proof}
Denote the first row of $[\mathbf{G}]$ and $[\bar{\mathbf{G}}]$ by
$(g_{0},g_{1},\ldots,g_{n-1})$ and
$(\bar{g}_{0},\bar{g}_{1},\ldots,\bar{g}_{n-1})$, respectively.
Thus, $g_{j}=\bar{g}_{j}=0$ for $n-k+1\le j\le n-1$. Since
$\alpha$ has order $n$ by definition, the element of
$[\mathbf{G}]$ and $[\bar{\mathbf{G}}]$ located at the
$(s+1)^{th}$ row and $(t+1)^{th}$ column, where $0\le s,t\le n-1$,
are $g_{t-s\bmod n}$ and $\bar{g}_{t-s\bmod n}=g_{t-s\bmod
n}\alpha^{(b-1)t}/\alpha^{(b-1)s}$, respectively.

Hence, $[\mathbf{\bar{G}}]$ may be obtained by multiplying the
$(t+1)^{th}$ column of $[\mathbf{G}]$ by $\alpha^{(b-1)t}$ and
dividing the $(s+1)^{th}$ row of the resultant matrix by
$\alpha^{(b-1)s}$. In matrix form, $[\bar{\mathbf{G}}]=
\mathbf{W}^{-1}\times[\mathbf{G}]\times \mathbf{W}$.
\end{proof}

Let $\mathbf{F}$ and $\mathbf{F}^{-1}$ denote the $n$-point GFFT
and inverse GFFT matrices over $\GF(q)$, ie.
\begin{equation}\label{gfft}
    \mathbf{F}=\left(%
\begin{array}{cccc}
  1 & 1 & \cdots & 1 \\
  1 & \alpha & \cdots & \alpha^{n-1} \\
  \vdots & \vdots & \ddots & \vdots \\
  1 & \alpha^{n-1} & \cdots & \alpha^{(n-1)(n-1)} \\
\end{array}%
\right),
\end{equation}
\begin{equation}\label{igfft}
    \mathbf{F}^{-1}=\left(%
\begin{array}{cccc}
  1 & 1 & \cdots & 1 \\
  1 & \alpha^{-1} & \cdots & \alpha^{-(n-1)} \\
  \vdots & \vdots & \ddots & \vdots \\
  1 & \alpha^{-(n-1)} & \cdots & \alpha^{-(n-1)(n-1)} \\
\end{array}%
\right).
\end{equation}
Lemma \ref{l3} gives a decomposition of $[\bar{\mathbf{G}}]$ in
terms of $\mathbf{F}$ and $\mathbf{F}^{-1}$.
\begin{lemma}\label{l3}
$[\bar{\mathbf{G}}]=\mathbf{F}^{-1}\times \mathbf{D} \times
\mathbf{F}$ where $\mathbf{D}$ is an $n\times n$ diagonal matrix
such that its main diagonal is the inverse GFFT of the first row
of $[\bar{\mathbf{G}}]$.
\end{lemma}
\begin{proof}
Let the inverse GFFT of the first row of $\bar{\mathbf{[G]}}$ be
$\mathbf{G}(1)=(G_{0},G_{1},\ldots,G_{n-1})$. Since the
$(i+1)^{th}$ row of $[\bar{\mathbf{G}}]$ is the right cyclic shift
of the first row of $[\bar{\mathbf{G}}]$ by $i$ positions, it
follows from the modulation property of GFFT\cite[Figure
6.1]{blahut:alg} that the inverse GFFT of the $(i+1)^{th}$ row of
$\bar{\mathbf{[G]}}$ is
$$\mathbf{G}(i+1)=(G_{0},G_{1}/\alpha^{i},G_{2}/\alpha^{2i},\ldots,G_{n-1}/\alpha^{(n-1)i}).$$
Consequently, the inverse GFFT of the rows of $[\bar{\mathbf{G}}]$
in matrix form may be expressed as
\begin{eqnarray}\label{igfftg}
 \nonumber [\mathbf{\bar{G}}]\times\mathbf{F}^{-1} &=& \left(%
\begin{array}{ccccc}
  G_{0} & G_{1} & G_{2} & \cdots & G_{n-1} \\
  G_{0} & G_{1}/\alpha & G_{2}/\alpha^{2} & \cdots & G_{n-1}/\alpha^{n-1} \\
  \vdots & \vdots & \vdots & \ddots & \vdots \\
  G_{0} & G_{1}/\alpha^{n-1} & G_{2}/\alpha^{2(n-1)} & \cdots & G_{n-1}/\alpha^{(n-1)(n-1)} \\
\end{array}%
\right) \\
  \nonumber &=& \left(%
\begin{array}{ccccc}
  1 & 1 & 1 & \cdots & 1 \\
  1 & \alpha^{-1} & \alpha^{-2} & \cdots & \alpha^{-(n-1)} \\
  \vdots & \vdots & \vdots & \ddots & \vdots \\
  1 & \alpha^{-(n-1)} & \alpha^{-2(n-1)} & \cdots & \alpha^{-(n-1)(n-1)} \\
\end{array}%
\right)\left(%
\begin{array}{cccc}
  G_{0} & 0  & \cdots & 0 \\
  0 & G_{1}  & \cdots & 0 \\
  0 & 0  & \ddots & 0 \\
  0 & 0  & \cdots & G_{n-1} \\
\end{array}%
\right)\\
&=&\mathbf{F}^{-1}\times\left(%
\begin{array}{cccc}
  G_{0} & 0  & \cdots & 0 \\
  0 & G_{1}  & \cdots & 0 \\
  \vdots & \vdots  & \ddots & \vdots \\
  0 & 0  & \cdots & G_{n-1} \\
\end{array}%
\right)=\mathbf{F}^{-1}\times\mathbf{D}
\end{eqnarray}
The lemma is now immediate.
\end{proof}

Since $\bar{g}(x)$ has zeros
$\alpha,\alpha^{2},\ldots,\alpha^{n-k}$, it follows from property
$2$ of \cite[Theorem 8-13]{Stephen:code} that the last $n-k$
elements of $\mathbf{G}(1)$ and in turn, the last $n-k$ elements
in the main diagonal of $\mathbf{D}$, are all zero. Moreover,
since $\bar{g}_{i}=g_{i}\alpha^{(b-1)i}$, by the translation
property of the GFFT\cite[Figure 6.1]{blahut:alg}, the inverse
GFFT $(G_{0},G_{1},\ldots,G_{k-1},0,\ldots,0)$ of the $n$-tuple
$(\bar{g}_{0},\bar{g}_{1},\ldots,\bar{g}_{n-k},0,\ldots,0)$ is the
right cyclic shift of the inverse GFFT of the $n$-tuple
$(g_{0},g_{1},\ldots,g_{n-k},0,\ldots,0)$ by $b-1$ positions.
\section{THE KEY RESULT}\label{sectkey}
In this section, we present a method for retrieving from the
output list of an interpolation based list decoder, messages
encoded as codewords of the RS code $\mathcal{C}$ via a generator
matrix $\mathbf{G}_{a}$ given by $\mathbf{G}_{a}=\mathbf{A}\times
\mathbf{G}$ for some $k\times k$ basis transformation matrix
$\mathbf{A}$. We assume that the decoder operates on the premise
that messages are encoded as codewords of $\bar{\mathcal{C}}$ via
the evaluation map.

Let $\tilde{\mathbf{A}}= (\begin{array}{cc}
  \mathbf{A} & \mathbf{0} \\
\end{array})$ be a $k\times n$  matrix where $\mathbf{0}$ is a
$k\times (n-k)$ all-zero matrix. In addition, let
$\mathbf{U}_{i\times j}$ denote the $i\times j$ upper-left
submatrix of $\mathbf{U}$. Further, since the evaluation map may
be interpreted as the $n$-point GFFT over $\GF(q)$, the relation
between a codeword $\bar{\mathbf{c}}$ of  $\bar{\mathcal{C}}$ and
its preimage $(f_{0},f_{1},\ldots,f_{k-1})\in GF(q)^{k}$ under
this map may be expressed as $\bar{\mathbf{c}}=\mathbf{f}\times
\mathbf{F}$ where
$\mathbf{f}=(f_{0},f_{1},\ldots,f_{k-1},0,\ldots,0)$. We can now
prove our main result, ie., Theorem \ref{t1}.
\begin{theorem}\label{t1}
Let $\mathbf{m}\in \GF(q)^k$ be encoded as $\mathbf{c}\in
\mathcal{C}$ via the generator matrix $\mathbf{G}_{a}$. Then
$\mathbf{m}^{T}=(\mathbf{A}^{T})^{-1}\times(\mathbf{W}_{k\times
k})^{-1}\times (\mathbf{F}^{-1}_{k\times
k})^{-1}\times(\mathbf{D}_{k\times k})^{-1}\times
(\mathbf{f}_{1\times k})^{T}$.
\end{theorem}
\begin{proof}
By Lemmas \ref{l1} to \ref{l3},
\begin{eqnarray}\label{ft1}
  \nonumber \bar{\mathbf{c}}&=& \mathbf{c}\times \mathbf{W}=\mathbf{m}\times \mathbf{G}_{a}\times \mathbf{W}=\mathbf{m}\times\mathbf{A}\times\mathbf{G}\times \mathbf{W} \\
  \nonumber &= &\mathbf{m}\times\tilde{\mathbf{A}}\times [\mathbf{G}]\times \mathbf{W}=\mathbf{m}\times\tilde{\mathbf{A}}\times \mathbf{W}\times[\mathbf{\bar{G}}] \\
  &=&\mathbf{m}\times\tilde{\mathbf{A}}\times \mathbf{W}\times \mathbf{F}^{-1}\times
  \mathbf{D}\times \mathbf{F}.
\end{eqnarray}
Since $\bar{\mathbf{c}}=\mathbf{f}\times \mathbf{F}$, we have
$\mathbf{f}=\mathbf{m}\times\tilde{\mathbf{A}}\times
\mathbf{W}\times \mathbf{F}^{-1}\times
  \mathbf{D}$. Moreover, since $\mathbf{F}$, $\mathbf{F}^{-1}$, $\mathbf{D}$ and $\mathbf{W}$ are
  symmetric,
\begin{equation}\label{ft2}
    \mathbf{f}^{T}=\mathbf{D}\times \mathbf{F}^{-1}\times \mathbf{W}\times\tilde{\mathbf{A}}^{T}\times \mathbf{m}^{T}.
\end{equation}
Since the last $n-k$ elements in the main diagonal of $\mathbf{D}$
as well as in the column vector $\tilde{\mathbf{A}}^{T}\times
\mathbf{m}^{T}$ are all zero, the last $n-k$ constraints in
(\ref{ft2}) vanish and so (\ref{ft2}) may be reduced to
$(\mathbf{f}^{T})_{k\times 1}=\mathbf{D}_{k\times k}\times
(\mathbf{F}^{-1}\times \mathbf{W})_{k\times
k}\times(\tilde{\mathbf{A}}^{T}\times \mathbf{m}^{T})_{k\times
1}$. Because $\mathbf{W}$ is diagonal, $(\mathbf{F}^{-1}\times
\mathbf{W})_{k\times k}=\mathbf{F}^{-1}_{k\times k}\times
\mathbf{W}_{k\times k}$. Moreover, $(\tilde{\mathbf{A}}^{T}\times
\mathbf{m}^{T})_{k\times 1}= \mathbf{A}^{T}\times \mathbf{m}^{T}$
and since $\mathbf{F}^{-1}_{k\times k}$, $\mathbf{D}_{k\times k}$,
$\mathbf{W}_{k\times k}$ and $\mathbf{A}$ are all invertible, it
follows that
$\mathbf{m}^{T}=(\mathbf{A}^{T})^{-1}\times(\mathbf{W}_{k\times
k})^{-1}\times (\mathbf{F}^{-1}_{k\times
k})^{-1}\times(\mathbf{D}_{k\times k})^{-1}\times
(\mathbf{f}_{1\times k})^{T}$.
\end{proof}

Interpreting $\mathbf{f}_{1\times k}$ and $\mathbf{m}$ in Theorem \ref{t1} as
an element of the output list and the desired data block,
respectively, leads to Algorithm \ref{algo1} below which
summarizes the key steps involved to recover any
generator-matrix-based RS coded data from the output list of an
interpolation-based list decoder.
\begin{algorithm}{\textbf{ }}\label{algo1}\\
Input: The zeros
$(\alpha^{b},\alpha^{b+1},\ldots,\alpha^{n-k-1+b})$ of
$\mathcal{C}$ and its generator matrix $\mathbf{G}_{a}$.\\
Output: The desired messages corresponding to the elements of the
output list.\\
Precomputation (to be performed only once): \begin{enumerate}
\renewcommand{\labelenumi}{\roman{enumi}.}
    \item  Compute $g(x)=\prod^{n-k-1}_{i=0}(x-\alpha^{b+i})=\sum^{n-k}_{i=0}g_{i}x^{i}$ and
    construct the matrix $\mathbf{G}$ for which the
    $(i+1)^{th}$ row, $0\le i\le n-1$, is the right cyclic shift of the $n$-tuple $(g_{0},g_{1},\ldots,
g_{n-k},0,\ldots,0)$ by $i$ positions.
    \item Find $\mathbf{A}$ such that $\mathbf{G}_{a}=\mathbf{A}\times
    \mathbf{G}$ and $(\mathbf{A}^{T})^{-1}$. (Note: The matrix $\mathbf{A}$ can be easily found using standard techniques in linear algebra since $\mathbf{G}$ is in row echelon form.)
    \item Compute the inverse GFFT of the $n$-tuple $(g_{0},g_{1},\ldots,
g_{n-k},0,\ldots,0)$. Then right cyclic shift the resultant vector
by $b-1$ positions to obtain
$(G_{0},G_{1},\ldots,G_{k-1},0,\ldots,0)$.
    \item Set $(D_{k\times
    k})^{-1}=\mbox{diag}(G_{0}^{-1},G_{1}^{-1},\ldots,G_{k-1}^{-1})$ and $(\mathbf{W}_{k\times
k})^{-1}=\mbox{diag}(1,\alpha^{-(b-1)},\alpha^{-2(b-1)},\ldots,\\\alpha^{-(k-1)(b-1)})$.
    \item Compute $(\mathbf{F}^{-1}_{k\times k})^{-1}$ and $\mathbf{B}=(\mathbf{A}^{T})^{-1}\times(\mathbf{W}_{k\times k})^{-1}\times(\mathbf{F}^{-1}_{k\times
k})^{-1} \times (D_{k\times
    k})^{-1}$. (Note: Since $\mathbf{F}^{-1}_{k\times k}$ is symmetric, its inverse can be
computed by eigenvalue decomposition.)
\end{enumerate}
List decoding $\&$ message recovery:
\begin{enumerate}
\renewcommand{\labelenumi}{\arabic{enumi}.}
    \item\label{s1} Compute $\mathbf{\bar{r}}=\mathbf{r}\times \mathbf{W}=(r_{0},r_{1}\alpha^{(b-1)},\ldots,r_{n-1}\alpha^{(n-1)(b-1)})$ where $\mathbf{r}$ is the hard-decision received vector.
    \item\label{s2} List decode $\mathbf{\bar{r}}$.
    \item\label{s3} If the output list is not empty, then for each element $\mathbf{f}_{1\times
    k}$ in this list, return $\mathbf{B}\times (\mathbf{f}^{T})_{k\times
    1}$.
\end{enumerate}
\end{algorithm}

A few remarks are in order. First, since the average list size is
typically very close to unity
\cite{mceliece:listsize}\cite{kotter:vardy}, Steps \ref{s1}) $\&$
\ref{s3}) will incur close to $k^{2}+n-1$ $\GF(q)$-multiplications
on average. Thus, for code rates of practical interest, an average
computational overhead of $O(k^{2})$ multiplications is introduced
on top of the computations incurred by Step \ref{s2}). Second, if
$b=1$, $\mathbf{W}$ reduces to an identity matrix such that
$\mathbf{W}_{k\times k}$ may be omitted in the computation of
$\mathbf{B}$. Finally, if messages were originally encoded by
multiplication by $g(x)$, then $\mathbf{G}_{a}=\mathbf{G}$ in this
case with $\mathbf{A}$ reducing to an identity matrix.
\begin{example}
Let $\mathcal{C}$ be a $(7,4)$ RS code over $\GF(8)$ with zeros
$\alpha^{2},\alpha^{3},\alpha^{4}$. Its generator matrix is
$$\mathbf{G}_{a}=\left(%
\begin{array}{ccccccc}
  \alpha^{5} & \alpha & \alpha^{3} & \alpha & \alpha^{3} & \alpha^{2} & \alpha \\
  \alpha^{6} & 0 & \alpha^{4} & \alpha^{3} & \alpha^{6} & 1 & \alpha^{2} \\
  \alpha^{6} & \alpha^{2} & \alpha^{2} & \alpha^{2} & 0 & \alpha^{5} & \alpha^{6} \\
  \alpha^{4} & \alpha^{6} & \alpha^{3} & \alpha^{2} & 1 & 0 & \alpha \\
\end{array}%
\right).$$  Following Algorithm \ref{algo1}, we obtain
$$(\mathbf{A}^{T})^{-1}=\left(%
\begin{array}{cccc}
  \alpha^{2} & 1 & \alpha^{2} & 0 \\
  \alpha^{2} & \alpha & \alpha^{2} & \alpha \\
  \alpha^{3} & \alpha^{6} & \alpha^{5} & \alpha^{5} \\
  \alpha^{6} & \alpha^{3} & \alpha^{2} & \alpha \\
\end{array}%
\right).$$ Applying the inverse GFFT to
$(g_{0},g_{1},g_{2},g_{3},0,0,0)=(\alpha^{2},\alpha^{3},1,1,0,0,0)$
and right cyclic shifting the resulting vector by $1$ position
yields
$(G_{0},G_{1},G_{2},G_{3},0,0,0)=(\alpha^{6},\alpha^{5},1,\alpha^{5},0,0,0)$
and so $(\mathbf{D}_{4\times
4})^{-1}=\mbox{diag}(\alpha,\alpha^{2},1,\alpha^{2})$. Now,
$(\mathbf{W}_{4\times
4})^{-1}=\mbox{diag}(1,\alpha^{6},\alpha^{5},\alpha^{4})$ and
$$(\mathbf{F}^{-1}_{4\times 4})^{-1}=\left(%
\begin{array}{cccc}
  \alpha^{4} & \alpha^{3} & \alpha^{5} & \alpha^{3} \\
  \alpha^{3} & 1 & 0 & \alpha \\
  \alpha^{5} & 0 & \alpha^{3} & \alpha^{2} \\
  \alpha^{3} & \alpha & \alpha^{2} & \alpha^{6} \\
\end{array}%
\right).$$ Hence,
\begin{equation}\label{mb}
    \mathbf{B} = (\mathbf{A}^{T})^{-1}\times
(\mathbf{W}_{k\times k})^{-1}\times(\mathbf{F}^{-1}_{k\times
k})^{-1} \times (D_{k\times
    k})^{-1}=\left(%
\begin{array}{cccc}
  \alpha^{5} & \alpha^{3} & \alpha & \alpha \\
  \alpha^{4} & \alpha^{5} & \alpha^{3} & 1 \\
  \alpha^{5} & \alpha^{2} & 1 & \alpha \\
  \alpha & \alpha & \alpha^{2} & \alpha \\
\end{array}%
\right).
\end{equation}
Suppose the codeword
$\mathbf{c}=(\alpha^{2},0,\alpha,0,0,\alpha^{3},\alpha^{6})$ is
transmitted and received as $\mathbf{r}$. If
 list decoding the vector $\bar{\mathbf{r}}=\mathbf{r}\times \mathbf{W}$ is successful,
$\mathbf{f}=(\alpha,0,\alpha^{5},1,0,0,0)$ will be in the output
list. We can recover the original message
$\mathbf{m}^{T}=\mathbf{B}\times \mathbf{f}^{T}_{4\times
1}=(\alpha^{3},\alpha^{2},0,\alpha^{5})^{T}$. It can be verified
that $\mathbf{c}=\mathbf{m}\times \mathbf{G}_{a}$.
\end{example}
\section{CONCLUSION}
To summarize, we have established a relationship between codewords resulting from generator-matrix-based encoding, and codewords obtained via the evaluation map. We have further derived from this relationship, an algorithm for recovering generator-matrix-based coded data under interpolation-based list decoding.

\end{document}